\documentclass[11pt,a4paper]{article}
\begin{document}

\title{ integral equations for the spin-weighted spheroidal wave function }
\author{Guihua Tian\\
School of Science, Beijing University \\
of Posts And Telecommunications. Beijing100876, China.\\Academy of
Mathematics and Systems Science,\\ Chinese Academy of
Sciences,(CAS) Beijing, 100080, P.R. China.}
\date{October 18, 2004}
\maketitle

\begin{abstract}
Integral equations for the spin-weighted spheroidal wave functions
is given . For the prolate spheroidal wave function with
 $ m=0 $, there exists the integral equation whose kernel is $
\frac{\sin x}{x}$ , and the sinc  function kernel $ \frac{\sin
x}{x}$
 is of great mathematical significance. In the paper, we
also extend the similar sinc function kernel $ \frac{\sin x}{x}$
to the case $m\neq 0 $ and $s\neq 0$, which interestingly turn out
as some kind of Hankel transformation.

\end{abstract}

Perturbation to black-hole is important not only in principle but
also astrophysically in reality. Teukolsky obtain separable
equations of scalar, electromagnetic and gravitational fields'
perturbation to the kerr black-hole\cite{1}-\cite{2}. The
separated angular equation are called the  spin-weighted
spheroidal wave equations, and  are extension of the ordinary
spherical harmonics equation. The spin-weighted spheroidal wave
functions  are not only connected to stability of kerr black hole
in general relativity, but also to many other physical
problems\cite{13}.  They are studied in many papers
\cite{1}-\cite{10}, though not by the method of the integral
equation.

It is well-known that the method of the integral equation play an
important role in mathematical physics. Many important theoretical
and mathematical problems are solved mainly by the method of the
integral equation. They are also indispensable for numerical
computation. Therefore, it is useful for one to find the integral
equation of a given differential equation, or equivalently, to
find the  kernel of the integral equation. In this paper, we
mainly give an integral equation of these spin-weighted spheroidal
wave functions. When reducing to the prolate spheroidal wave
function, there exists the integral equation whose kernel is $
\frac{\sin x}{x}$ when $ m=0 $, the $ \frac{\sin x}{x}$ kernel is
of great mathematical significance\cite{11}, \cite{12}. In the
paper, we also extend the similar $ \frac{\sin x}{x}$ kernel to
the case $m\neq 0 $ and $s\neq 0$, which interestingly turn out as
some kind of Hankel transformation.

The spin-weighted spheroidal wave functions $Z_{slm}$ connect with
$S_{slm}$ by the relation
\begin{equation}
Z_{slm}=\left[\frac{2l+1}{4\pi}
\frac{(l-m)!}{(l+m)!}\right]S_{slm}(-a\omega, \cos \theta)
e^{im\phi} \label{1}
\end{equation}

Where $S_{slm}$ satisfy the angular part of the perturbation wave
equation
\begin{equation}
\left[\frac{1}{\sin \theta}\frac{d}{d\theta}(\sin \theta
\frac{d}{d \theta})+s+ a^{2}\omega ^2\cos ^2 \theta   -2as\omega
\cos \theta -\frac{(m+s\cos \theta)^{2}}{\sin ^2
\theta}\right]S_{slm}=-A_{slm}S_{slm}\label{2}
\end{equation}
for eigenvalue $A_{slm}$.

The quantities $a,s,\omega,m$ are respectively the specific
angular momentum of kerr black hole, the spin, frequency and
magnetic momentum of the perturbation field $\psi$
\cite{1}-\cite{3}.The equation (\ref {2}) reduces to spin-weighted
associated Legendre-polynomials when $a\omega=0$; to associated
Legendre-polynomials when $a\omega=0$ and $s=0$, to spheroidal
function (or oblate spheroidal function) when $s=0$\cite{9}. In
the following, we study the integral equation of the eq. (\ref
{2}). Letting $x=\cos \theta$, and for convenience, denoting
$A_{slm},S_{slm}$ and $a^{2}\omega^{2}$ by $A,S,b^2$ respectively,
the eq. (\ref{2}) now becomes\cite{1}-\cite{8}
\begin{equation}
\frac{d}{dx}\left[(1-x^2) \frac{d}{d x}\right]+\left[A+s+ b^2x^2
 -2bsx-\frac{(m+sx)^{2}}{1-x^2} \right]S=0.\label{3}
\end{equation}
In the following, $m+s\geq 0$ and $m-s\geq 0$ is supposed.

Just as in the book of Flammer's, define operator $L_x$ as\cite{9}
\begin{equation}
L_x=\frac{\partial}{\partial x}\left[(1-x^2)
\frac{\partial}{\partial x}\right]+\left[s+ b^2x^2
 -2bs x-\frac{(m+sx)^{2}}{1-x^2} \right],\label{41}
\end{equation}
then , the spin-weighted spheroidal wave function $S$ satisfies
the following equation
\begin{equation}
\left[L_x+A\right]S=0\label{4}.
\end{equation}
Let $K(x,y)$have the first and the second continuous derivatives
with respect to $x$ and $y$, and satisfy the equations
\begin{equation}
\left[L_x-L_y\right]K(x,y)=0\label{5}
\end{equation}
and
\begin{equation}
\left[(1-x^2)S(x)\frac{\partial K(x,y)}{\partial x
}-(1-x^2)K(x,y)\frac{d S(x)}{dx }\right]_{x=-1}^{x=1}=0,\label{6}
\end{equation}
then, we have
\begin{equation}
\left[L_y+A\right]\bar{S}(y)=0,\label{7}
\end{equation}
where $\bar{S}$ is defined by
\begin{equation}
\bar{S}(y)=\int_{-1}^{+1}K(x,y)S(x)dx.\label{8}
\end{equation}
The above conclusion is easy to prove, that is,
\begin{eqnarray}
L_{y} \bar{S} &=& \int_{-1}^{+1}\left[L_yK(x,y)\right]S(x)dx\nonumber\\
&=& \int_{-1}^{+1}\left[L_xK(x,y)\right]S(x)dx \nonumber\\
&=& \int_{-1}^{+1}\frac{\partial }{\partial x
}\left[(1-x^2)S(x)\frac{\partial K(x,y)}{\partial x
}-(1-x^2)K(x,y)\frac{d S(x)}{dx
}\right]dx\nonumber\\&+&\int_{-1}^{+1}K(x,y)L_xS(x)dx\nonumber\\
&=&-A\int_{-1}^{+1}K(x,y)S(x)dx=-A\bar{S}(y),
\end{eqnarray}
where eqs.(\ref{4})-(\ref{6}),(\ref{8}) have been used. The
equation (\ref{7}) means $\bar{S}(x)$ satisfies the same equation
as the eigenfunction $S(x)$ of the eq.(\ref{3})(equivalently the
eq.(\ref{4})). There is only one independent eigenfunction for the
equation (\ref{3}), so, $\bar{S}(x)$ is the same as $S(x)$ up to a
constant factor, that is, $S(x)\propto \bar{S}(x)$. Therefore, the
eq.(\ref{8}) is really the integral equation form of the
eq.(\ref{3}), where the kernel $K(x,y)$ must satisfies the
equations (\ref{5}),(\ref{6}).

We really find one such kernel $K(x,y)$ which reads
\begin{eqnarray}
K(x,y)=C_1(1-x)^{\frac12(m+s)}(1+x)^{\frac12(m-s)}(1-y)^{\frac12(m+s)}(1+y)^{\frac12(m-s)}e^{bxy}.\label{9}
\end{eqnarray}
Similarly,
\begin{eqnarray}
\bar{K}(x,y)=C_2(1-x)^{\frac12(m+s)}(1+x)^{\frac12(m-s)}(1-y)^{\frac12(m+s)}(1+y)^{\frac12(m-s)}e^{-bxy}\label{10}
\end{eqnarray}
also satisfies the eqs.(\ref{5}),(\ref{6}),and is another kernel
for the spin-weighted spheroidal wave function's integral
equation. These kernels reduce to the familiar form for the oblate
spheroidal wave function when $s=0$; further, to the familiar form
of the prolate spheroidal wave function when the parameter $b$
becomes imaginary, that is,
\begin{eqnarray}
K(x,y)&=&C_1(1-x^2)^{\frac12m}(1-y^2)^{\frac12m}e^{icxy}\label{11}\\
\bar{K}(x,y)&=&C_2(1-x^2)^{\frac12m}(1-y^2)^{\frac12m}e^{-icxy}\label{12}
\end{eqnarray}
where $b=ic$, $c\in R$.

In the following, we study the integral equation  of the prolate
spheroidal wave function, which corresponds the equation (\ref{3})
when the spin $s$ is zero and the parameter $b$ be imaginary
($b=ic$, $c\in R$). When $m=0$ , there is the kernel of the form
$\frac{\sin x}{x}$ for the prolate spheroidal wave function. The
kernel $\frac{\sin x}{x}$ is of great significance in the
communication science. D. Slepian   extend the form to the case
$m\neq 0$ by the Bessel function\cite{12}. Here, we make another
extension to the case $m\neq 0$.

The prolate spheroidal wave equation is
\begin{equation}
\frac{d}{dx}\left[(1-x^2) \frac{d}{d x}\right]+\left[A-c^2x^2-
 \frac{m^{2}}{1-x^2} \right]S_m=0.\label{13}
\end{equation}
When $m=0$, the integral equation of the kernel $\frac{\sin x}{x}$
is
\begin{equation}
S_0(y)=\int_{-1}^{+1}\frac{\sin (x-y)c}{\pi
(x-y)}S_0(x)dx\label{14}
\end{equation}
(see the references \cite{11},\cite{12} and the references cited
there). When $m\neq 0 $, $S_m(x)$ has the integral equations whose
kernel are of the forms of eq.(\ref{11}) and eq.(\ref{12}). From
these equations, we get
\begin{equation}
S_m(y)=\int_{-1}^{+1}(1-x^2)^{\frac12m}(1-y^2)^{\frac12m}I_m(x,y)S_m(x)dx,\label{15}
\end{equation}
where $I_m(x,y)$ is defined by
\begin{equation}
I_m(x,y)=\int_{-1}^{+1}(1-t^2)^{m}e^{ic(x-y)t}dt.\label{16}
\end{equation}
When $m=0$,
\begin{equation}
I_0(x,y)=2\frac{\sin (x-y)c}{c(x-y)}\label{17}
\end{equation}
is reduced to the kernel $\frac{\sin x}{x}$; When $m=1$,
\begin{equation}
I_1(x,y)=\frac{4\sin (x-y)c-4c(x-y)\cos
(x-y)c}{c^3(x-y)^3};\label{18}
\end{equation}
When $m\geq 2$, $I_m$ satisfy the following relation
\begin{equation}
I_m(x,y)=\frac{2m(2m-1)I_{m-1}(x,y)}{c^2(x-y)^2}-\frac{4m(m-1)I_{m-2}(x,y)}{c^2(x-y)^2}.\label{19}
\end{equation}
This complete the extension.

The function $I_m(x,y)$ are really connected with Bessel
functions, that is, by eq.(\ref{16})
\begin{eqnarray}
I_m(x,y)&=&\int_{-1}^{+1}(1-t^2)^{m}\cos\left(c(x-y)t\right)dt+i\int_{-1}^{+1}(1-t^2)^{m}\sin\left(c(x-y)t\right)dt\\
&=&\int_{-1}^{+1}(1-t^2)^{m}\cos\left(c(x-y)t\right)dt,\label{24}
\end{eqnarray}
from reference\cite{14}, it is easy to get
\begin{eqnarray}
J_n(z)&=&\int_{-1}^{+1}(1-t^2)^{n+\frac12}\cos\left(zt)\right)dt;\label{25}
\end{eqnarray}
therefore, we obtain
\begin{equation}
I_m(x,y)=\frac{2^{m+\frac12}\Gamma(m+1)\Gamma(\frac12)}{\left[c(x-y)\right]^{m+\frac12}}J_{m+\frac12}\left(c(x-y)\right).\label{26}
\end{equation}
Then, the eq.(\ref{15}) could be written as
\begin{equation}
S_m(y)=\int_{-1}^{+1}(1-x^2)^{\frac12m}(1-y^2)^{\frac12m}S_m(x)
\frac{2^{m+\frac12}\Gamma(m+1)\Gamma(\frac12)}{\left[c(x-y)\right]^{m+\frac12}}J_{m+\frac12}\left(c(x-y)\right)dx,\label{27}
\end{equation}
which is kind of Hankel transformation on finite interval. By
Hankel transformation, the equation (\ref{27}) contains many
useful information, which will be our further studies. In
completing the extension to $m\neq 0$,we realize that our results
are similar to that of D. Slepian's\cite{12}, or that of
Flammer's\cite{9}, though deduced by different method.

It is natural to make the integral equation of the form
(eq.(\ref{27})) extend to the spin-weighted spheroidal wave
function with $s\neq 0$. But direct calculation shows that this
work is not easy. From the kernels $K(x,y)$ and $\bar{K}(x,y)$, we
only could obtain the following integral equation:
\begin{equation}
S_m(y)=\int_{-1}^{+1}(1-x)^{\frac12(m+s)}(1+x)^{\frac12(m-s)}(1-y)^{\frac12(m+s)}(1+y)^{\frac12(m-s)}
\bar{I}_{m,s}(x,y)S_m(x)dx,\label{28}
\end{equation}
where $\bar{I}_{m,s}(x,y)$ is defined by
\begin{equation}
\bar{I}_{m,s}(x,y)=\int_{-1}^{+1}(1-t)^{m+s}(1+t)^{m-s}e^{ic(x-y)t}dt
\label{29}
\end{equation}
and $\ b=-ic$. By eq.(\ref{29}),it is easy to see that
$\bar{I}_{m,s}(x,y)$ reduce to $I_m(x,y)$(see
eqs.(\ref{16}),(\ref{26})). But the connection of
$\bar{I}_{m,s}(x,y)$ with the bessel function is not direct when
$s\neq 0$.

Interestingly, in order to obtain the similar results for $s\neq
0$ just as in the case $s=0$, we get by chance another two kernels
for the spin-weighted spheroidal wave functions with $s\neq 0$.
These two kernels are
\begin{eqnarray}
K_2(x,y)=C_1(1+x)^{\frac12(m+s)}(1-x)^{\frac12(m-s)}(1+y)^{\frac12(m+s)}(1-y)^{\frac12(m-s)}e^{bxy}\label{30}
\end{eqnarray}
and
\begin{eqnarray}
\bar{K}_2(x,y)=C_2(1+x)^{\frac12(m+s)}(1-x)^{\frac12(m-s)}(1+y)^{\frac12(m+s)}(1-y)^{\frac12(m-s)}e^{-bxy}.\label{31}
\end{eqnarray}
It is easy to check that the two kernels $K_2(x,y)$ and
$\bar{K}_2(x,y)$ really satisfy the two conditions
(eqs.(\ref{5}),(\ref{6})) for the kernels of the spin-weighted
spheroidal wave functions. From the four kernels $K(x,y)$,
$\bar{K}(x,y)$, $K_2(x,y)$ and $\bar{K}_2(x,y)$ of the
spin-weighted spheroidal wave functions, we could get the
following
\begin{equation}
S_m(y)=\int_{-1}^{+1}(1+x)^{\frac12(m+s)}(1-x)^{\frac12(m-s)}(1-y)^{\frac12(m+s)}(1+y)^{\frac12(m-s)}I_m(x,y)S_m(x)dx
,\label{32}
\end{equation}
and
\begin{equation}
S_m(y)=\int_{-1}^{+1}(1-x)^{\frac12(m+s)}(1+x)^{\frac12(m-s)}(1+y)^{\frac12(m+s)}(1-y)^{\frac12(m-s)}I_m(x,y)S_m(x)dx,
\label{33}
\end{equation}
where $I_m(x,y)$ is the same as that in the case $s=0$ (see
eqs.(\ref{16}), (\ref{26}))except $b=-ic$. In fact, $b=-ic$ means
that $I_m(x,y)$ is connected with the modified bessel function.
Obviously, these two forms are direct extension of the bessel
kernels to the spin-weighted spheroidal wave functions when $s\neq
0$. The equations (\ref{32}), (\ref{33})) are Hankel
transformation in nature, and will be our further study.

 Discussion: in the following, we rather make discussion of the form of eqs.(\ref{15})-(\ref{16}), as the
 function $I_m(x,y)$ is easily more similar with the function $I_0(x,y)$  in this form.
 From the recurrence relation (see eq.(\ref{19}))
of the functions $I_m(x,y)$, one could  deduce that $I_m(x,y)$ are
really a functional of the functions $I_0(x,y)$and $I_1(x,y)$.
$I_0(x,y)$ is the sinc function $\frac{\sin (x-y)c}{(x-y)c}$ , and
$I_1(x,y)$ has has some properties as that of the function
$I_0(x,y)$, such as

(1)\begin{eqnarray}\lim_{x\rightarrow
y}I_0(x,y)&=&\lim_{x\rightarrow y}\left[2\frac{\sin
(x-y)c}{c(x-y)}\right]=1\label{20} \\\lim_{x\rightarrow
y}I_1(x,y)&=&\lim_{x\rightarrow y}\left[\frac{4\sin
(x-y)c-4c(x-y)\cos (x-y)c}{c^3(x-y)^3}\right]\\\nonumber &=&
\frac43\label{21}
\end{eqnarray}

(2) as $|x-y|$ increasing , $I_0(x,y)$ oscillatorily  approaches
zero; similarly, $I_1(x,y)$ oscillatorily  approaches zero more
rapidly.

Therefore, it might not  be unreasonable to regard that the
$\frac{\sin (x-y)c}{(x-y)c}$ kernel extension to $m\neq 0$ case be
appropriate.

By eq.(\ref{19}), one might think that the function $I_m(x,y)$
could be unbounded as $x\rightarrow y$. This is not the case.
Really, from the definition of the functions $I_m(x,y)$ and the
properties of the fourier transformation, it is easy to see that
the functions $I_m(x,y)$ are bounded as $x\rightarrow y$, here we
give some examples:

Using eq.(\ref{19}) at $m=2$, one gets
\begin{eqnarray}
I_2(x,y)&=&\frac{12I_{1}(x,y)}{c^2(x-y)^2}-\frac{8I_{0}(x,y)}{c^2(x-y)^2}\nonumber\\
&=&\frac{48\left(4\sin (x-y)c-4c(x-y)\cos
(x-y)c\right)-16c^2(x-y)^2\sin (x-y)c}{c^5(x-y)^5}\nonumber\\
&&\lim_{x\rightarrow y}I_2(x,y)=\frac{15}{16} ;\label{22}
\end{eqnarray}
similarly , at $m=3$, one gets
\begin{eqnarray}
&&I_3(x,y)=\frac{30I_{2}(x,y)}{c^2(x-y)^2}-\frac{24I_{1}(x,y)}{c^2(x-y)^2}\nonumber\\
&=&\frac{1440\left(4\sin (x-y)c-4c(x-y)\cos
(x-y)c\right)-576c^2(x-y)^2\sin (x-y)c+96c^3(x-y)^3\cos
(x-y)c}{c^7(x-y)^7}\nonumber \\&&\lim_{x\rightarrow
y}I_3(x,y)=\frac{32}{35} ;\label{23}
\end{eqnarray}

Conclusion: in the paper, we get the kernels for the integral
equation of the spin-weighted spheroidal wave
functions(eqs.(\ref{9}),(\ref{10})); we also extend the
$sinc(x)$-like kernel to the case $m\neq 0$ and $s\neq
0$(eqs.(\ref{32}),(\ref{33})).

\section*{Acknowledgments}
Professor Yunkau Lau insists that his name be in acknowledge is
enough. We are supported in part by the National Science
Foundation of China under Grant No.10475013, No.10347151,
No.10373003, 10375008 and the post-doctor foundation of China.

\end{document}